# Separation and Quantification of Ionospheric Convection Sources: 1. A New Technique

J. P. Reistad[1], K. M. Laundal[1], N. Østgaard[1], A. Ohma[1], S. Haaland[1,2], K. Oksavik[1,3], and S. E. Milan[1,4]

[1]Birkeland Centre for Space Science, University of Bergen, Bergen, Norway, [2]Max Planck Institute for Solar System Research, Göttingen, Germany, [3]Arctic Geophysics, University Centre in Svalbard, Longyearbyen, Norway, [4]Department of Physics and Astronomy, University of Leicester, Leicester, UK

**Abstract** This paper describes a novel technique that allows separation and quantification of different sources of convection in the high-latitude ionosphere. To represent the ionospheric convection electric field, we use the Spherical Elementary Convection Systems representation. We demonstrate how this technique can separate and quantify the contributions from different magnetospheric source regions to the overall ionospheric convection pattern. The technique is in particular useful for distinguishing the contributions of high-latitude reconnection associated with lobe cells from the low-latitude reconnection associated with Dungey two-cell circulation. The results from the current paper are utilized in a companion paper (Reistad et al., 2019, https://doi.org/10.1029/2019JA026641) to quantify how the dipole tilt angle influences lobe convection cells. We also describe a relation bridging other representations of the ionospheric convection electric field or potential to the Spherical Elementary Convection Systems description, enabling a similar separation of convection sources from existing models.

## 1. Introduction

For decades, patterns of high-latitude plasma circulation have been inferred from measurements of the ionospheric convection velocity $\vec{v}$. In addition to in situ measurements from the ionosphere and magnetosphere, ground-based high-frequency radars, for example, the Super Dual Auroral Radar Network (SuperDARN; Chisham et al., 2007; Greenwald et al., 1995), have been an important tool in these investigations. In addition to describe climatological patterns of the high-latitude ionospheric convection (e.g., Cousins & Shepherd, 2010; Haaland et al., 2007; Heppner & Maynard, 1987; Greenwald et al., 1995; Thomas & Shepherd, 2018; Ruohoniemi & Greenwald, 2005; Weimer, 1995), the instantaneous ionospheric convection is routinely derived from SuperDARN by including statistical "fill-in-data" from an empirical model, known as the "map potential technique" (Ruohoniemi & Baker, 1998), representing a likely snapshot of the present large-scale ionospheric convection pattern.

In the reference frame of the radar, the $F$ region convection corresponds to a convection electric field $\vec{E} = -\vec{v} \times \vec{B}$, where $\vec{v}$ is the ionospheric convection velocity vector relative to the radar and $\vec{B}$ is the magnetic field at the measurement location. On time scales longer than a few 10 s, the convection electric field can be considered curl free (Milan, 2013). Then, the electric field can be written as the gradient of a scalar field, referred to as the electric potential $\Phi$. As $\Phi$ can be considered constant along the magnetic field lines when no parallel electric fields are present, it is convenient to represent $\Phi$ on a spherical shell as an expansion of spherical harmonic functions. This approach is by far the most common way to represent $\Phi$, where observations of the plasma drift are used to estimate the coefficients of the spherical harmonic functions describing $\Phi$.

Despite providing an efficient and powerful framework to reproduce a scalar field on a sphere, there are fundamental limitations of the spherical harmonic description. One inherent property is the repetitive nature of the spherical harmonic functions. A good data fit in one part of the sphere can affect the solution at a different location. Furthermore, the shape of the analysis area is either global or restricted to a spherical cap by applying boundary conditions (Haines, 1985), and the level of detail (degree and order) of the reconstruction is the same over the entire sphere or spherical cap.

This paper describes a different approach to represent the ionospheric convection. We outline a method that allows the ionospheric convection electric field $\vec{E}$ to be represented as the sum of the electric fields from a large number of *nodes*, each having their own electric field $\vec{E}_j$ associated with them. This is a specific application of the Spherical Elementary Current Systems (SECS) technique developed by Amm (1997) and Amm and Viljanen (1999). Hence, we also refer to the method as SECS, but for our application C refers to convection rather than current. In this paper we will show that one of the benefits of representing the convection electric field in this way is the ability to segment the ionospheric convection field into regions corresponding the magnetospheric source of the ionospheric convection. A particular application is the ability to isolate and quantify the contributions to the ionospheric convection from dayside and lobe reconnection. In our companion paper *Separation and quantification of ionospheric convection sources: 2. The dipole tilt angle influence on reverse convection cells during northward IMF*, referred to as Paper II, this method enables us to quantify the influence of the dipole tilt angle on the lobe cell circulation in the ionosphere.

Although segmenting the convection electric field into source regions is a new ability, the SECS representation of $\vec{E}$ has also other advantages compared to the spherical harmonic representation of $\Phi$. This technique allows $\vec{E}$ to be represented locally in limited regions of any shape on a spherical surface. Furthermore, the density of nodes can change across the analysis domain to compensate for data coverage or varying degree of structure of $\vec{E}$ at different locations.

Section 2 introduces the SECS representation of a vector field on a sphere, as developed by Amm (1997) and Amm and Viljanen (1999), and how this can be used to represent the convection electric field at high latitudes, highlighting the ability to separate and quantify the contributions from lobe and Dungey-type convection as driven from the solar wind-magnetosphere interactions. Although Amm et al. (2010) already presented how the SECS technique can be used to describe the ionospheric convection velocity field, we here repeat that description with one important difference, which is to represent the convection electric field. As will be shown in the next section, this difference is what enables us to do the segmenting of the ionospheric convection. Section 3 presents the physical interpretations that can be made from the SECS description of the convection electric field, as well as a comparison of the results from the SECS description with a recent spherical harmonic model of $\Phi$. Section 4 concludes the paper.

## 2. Using SECS to Represent the Ionospheric Convection Electric Field

By convection electric field we refer to the electric field due to the ionospheric plasma motion, as described by the Lorentz transformation $\vec{E} = -\vec{v} \times \vec{B}$, since we assume no electric field in the frame of the plasma motion. Hence, $\vec{v}$ is the bulk plasma velocity in the *F* region or above, relative to an observer, typically a ground station, and $\vec{B}$ is the magnetic field where $\vec{v}$ is measured.

### 2.1. Introduction to the SECS Representation of a Vector Field

According to the Helmholtz decomposition, any three-dimensional vector field, $\vec{u}$, can be represented by a superposition of a curl-free, $\vec{u}_{cf}$, and a divergence-free, $\vec{u}_{df}$, vector field: $\vec{u} = \vec{u}_{cf} + \vec{u}_{df}$. Based on this, Amm (1997) developed the functional form of a set of curl-free and divergence-free elementary vector fields that he showed would reconstruct any smooth vector field on a spherical surface by placing $n$ elementary field sources on the sphere, which we will refer to as nodes using the subscript *j*. In this description, the sum of the elementary fields from nodes with a curl-free field would represent $\vec{u}_{cf}$, and the sum of the elementary fields from nodes with a divergence-free field would represent $\vec{u}_{df}$:

$$\vec{u}_{cf} = \sum_{j=1}^{n} \vec{u}_{cf,j}$$
$$\vec{u}_{df} = \sum_{j=1}^{n} \vec{u}_{df,j} \tag{1}$$

From a set of requirements that we will return to later, Amm (1997) derived the functional form of these basis functions to be

$$\vec{u}_{cf,j}(\vec{r}_i) = \frac{A_j}{4\pi R} \cot(\theta_{ij}/2)\hat{\theta}_{ij}$$
$$\vec{u}_{df,j}(\vec{r}_i) = \frac{A_j}{4\pi R} \cot(\theta_{ij}/2)\hat{\phi}_{ij} \tag{2}$$

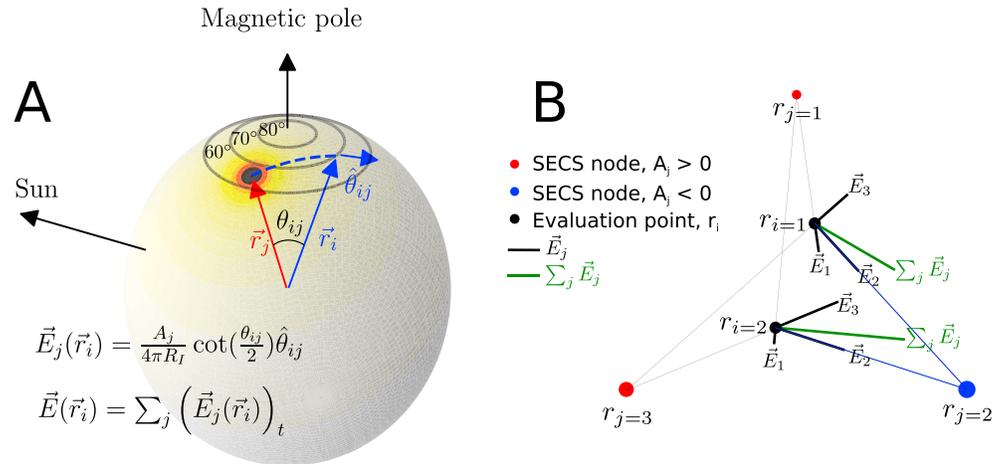

**Figure 1.** (a) Illustration of the elementary current function derived by Amm (1997) and the geometry involved in the Spherical Elementary Convection Systems (SECS) description. The global magnetic local time/magnetic latitude coordinate system is shown in black in a spherical shell representing the ionosphere at *F* region heights (300 km). The magnitude of the curl-free field from node *j*, located at $\vec{r}_j$ is shown in color. The unit vector of this field at position $\vec{r}_i$ is indicated as $\hat{\theta}_{ij}(\vec{r}_i)$. $\theta_{ij}$ is the angle between $\vec{r}_i$ and $\vec{r}_j$. (b) Three nodes, each representing a curl-free field, where the sign and magnitude of its associated amplitude $A_j$ is reflected by its color and size. At two arbitrary locations (black dots) the curl-free field from each node is shown as black vector pins, as well as its vector sum (green) representing the curl-free electric field described by the three nodes at that location.

Here, $\frac{A_j}{4\pi R}$ is scaling the strength of the elementary field, where $A_j$ is an amplitude for node *j*, placed on the spherical surface at radial distance *R*. A sketch illustrating the curl-free field from a single node and the relevant vectors involved is shown in Figure 1a. In Figure 1a we use the double subscript *ij* on quantities that depend both on the node location $\vec{r}_j$ and where the field is evaluated $\vec{r}_i$. Hence, the $\theta_{ij}$ argument is the angle from the position of the node, $\vec{r}_j$, to the position where the field is being evaluated, $\vec{r}_i$. In this description, equation (2) refers to a local coordinate system for each node. The strength of the field from node *j* is only dependent on this local polar angle $\theta_{ij}$. The unit vectors $\hat{\theta}_{ij}$ and $\hat{\phi}_{ij}$ also refer to this local frame, $\hat{\theta}_{ij}$ pointing away from the node, and $\hat{\phi}_{ij}$ in the perpendicular direction, along contours of constant $\theta_{ij}$. As mentioned, $\vec{u}_{cf}$ and $\vec{u}_{df}$ are the superposition of the fields from *n* nodes. To compute the superposition field from every node at a specific location ($\vec{r}_i$), the node fields need to be converted into a common coordinate system before the sum is calculated. Hence,

$$\vec{u}_{cf}(\vec{r}_i) = \sum_{j=1}^{n} \left( \frac{A_j}{4\pi R} \cot(\theta_{ij}/2)\hat{\theta}_{ij} \right)_t$$
$$\vec{u}_{df}(\vec{r}_i) = \sum_{j=1}^{n} \left( \frac{A_j}{4\pi R} \cot(\theta_{ij}/2)\hat{\phi}_{ij} \right)_t$$
(3)

where the subscript *t* highlights this coordinate conversion. Figure 1b illustrates the superposition field in an idealized case from three nodes each having a curl-free field, as expressed by equation (3a). The three nodes are shown as blue and red dots, where the color and its size indicate the sign and magnitude of its associated amplitude $A_j$. The curl-free field described by these three nodes is evaluated at two arbitrary locations (black dots) where the contributions from each node is shown (black vector pins) as well as the sum of all contributions at that location (green).

Amm and Viljanen (1999) used this approach to represent the equivalent currents in the ionosphere from measurements of their associated magnetic field perturbations detected on the ground. This approach they termed SECS, where the nodes could be interpreted as sources of divergence-free and curl-free currents, which they attributed to equivalent and field-aligned currents, respectively. In their application the nodes represent the source of a continuous divergence-free or curl-free current field (sheet current on a sphere) centered at the nodes. In the case of a curl-free node, this source is a field-aligned current. The amplitude $A_j$ in the elementary function (equation (2)), having units of amperes in their application, has the role of scaling the strength of the vector field from each node, normalized to the sphere radius *R*. Hence, the vector field itself represents sheet current density (A/m) in their application.

The SECS technique is a tool to represent any vector field on a sphere, where its curl-free and divergence-free components are represented as the superposition of curl-free or divergence-free fields from nodes placed on the surface of the same sphere. Given a distribution of *n* such SECS nodes, the task is to estimate all the corresponding *n* amplitudes that optimally fit observations of the same vector field. The SECS technique therefore has a wide range of applications. Amm et al. (2010) used the same technique and basis functions to describe the ionospheric convection *velocity* field. In their description, the ionospheric convection velocity field is purely divergence free (assume $\vec{B}$ is constant and the flow is incompressible). From benchmarking with a synthetic data set, they demonstrated that even when the line-of-sight (LOS) velocity data coverage was limited (25% of the analysis area), the SECS technique could reconstruct the velocity field with relative errors of less than ∼5% when the velocity field had a scale size of ∼100 km. Furthermore, in contrast to other ways of reconstructing the ionospheric convection velocity field, this technique requires in principle no a priori knowledge of the convection or boundary conditions but is only constrained by measurements and the assumption that the velocity field is incompressible.

### 2.2. Using SECS to Represent the Convection Electric Field

The present paper uses a slightly different approach to describe the ionospheric convection compared to Amm et al. (2010). Instead of representing the convection velocity field as a divergence-free field, we express the convection electric field as a curl-free field. This has a few advantages, as become evident when expanding the divergence-free condition using $\vec{E} = -\vec{v} \times \vec{B}$. If assuming $\nabla \times \vec{E} = 0$ in the steady state situation, and considering a magnetic field only due to sources inside the Earth, for example, as represented by the International Geomagnetic Reference Model, $\nabla \times \vec{B} = 0$. Then

$$\nabla \cdot \vec{v} = \vec{E} \cdot \nabla \left( \frac{1}{B^2} \right) \times \vec{B}. \tag{4}$$

Hence, inhomogeneities in the magnetic field along the convection path will contribute to a divergence of $\vec{v}$. Since we will combine measurements from different longitudes to make averages at specific magnetic local times (MLTs) and magnetic latitudes (MLATs), the representation in terms of a convection electric field is therefore beneficial when we want to relate the observed convection velocities to magnetic flux transport rates at different locations.

Since the convection electric field can be considered as curl free, we only need to use one of the elementary functions in equation (2) for the present application. The SECS electric field at location $\vec{r}_i$ in the global coordinate system (e.g., an MLT/MLAT system) then becomes

$$\vec{E}(\vec{r}_i) = \vec{E}_{cf}(\vec{r}_i) = \sum_{j=1}^{n} \left( \frac{A_j}{4\pi R_I} \cot(\theta_{ij}/2) \hat{\theta}_{ij} \right)_t. \tag{5}$$

Here, $\theta_{ij}$ is the angle between $\vec{r}_j$ and $\vec{r}_i$, as illustrated for one specific node in Figure 1a. $\hat{\theta}_{ij}$ is the corresponding unit vector of the curl-free elementary field of node *j* evaluated at $\vec{r}_i$, and the subscript *t* is to indicate that the contribution from each node need to be converted to a common coordinate system before the superposition field can be calculated. For this particular application of the SECS description, the SECS nodes must be placed at a minimum radial distance to ensure that $\vec{E} = -\vec{v} \times \vec{B}$. We use the Earth radius plus the height of the *F* region ionosphere, set to 300 km, referred to as $R_I$. Since the electric field has units of volts per meter, $A_j$ has units of volts in this description, or equivalently weber per second, highlighting its interpretations in terms of magnetic flux transport.

### 2.3. Electric Potential in the SECS Description

We here express the relationship between the SECS node amplitudes $A_j$ and $\Phi$, allowing a detailed comparison with the more standard representation of the ionospheric convection. The potential at a given point in the analysis domain is given by the sum of the potential from all electric field sources. As will be shown in the next subsection, a physical interpretation of the sources of the node electric field is indeed a charge distribution. Since $\vec{E} = -\nabla \Phi$, the potential is found by integrating equation (5). The potential at location $\vec{r}_i$ then becomes

$$\Phi(\vec{r}_i) = \sum_{j=1}^{n} \frac{-A_j}{2\pi} \ln \left( \sin(\theta_{ij}/2) \right). \tag{6}$$

Note that we have multiplied by $R_I$ as the integration of the elementary function is done along $\hat{\theta}_{ij}$ in a distance $R_I$ from origo, resulting in $\Phi$ having units of volts. As the potential is a scalar quantity, no coordinate transformation is needed to sum the contributions from each node.

### 2.4. Estimating the SECS Node Amplitudes $A_j$

Earlier we described how the set of amplitudes $A_j$ describes the electric field at an arbitrary location in the analysis domain, $\vec{r}_i$. Now we use the relationship from equation (5) to describe how $A_j$ can be estimated from observation of the electric field. Hence, when describing the inversion for $A_j$ in the following, $\vec{r}_i$ refer to the location of observation $i$.

Equation (5) shows that the observations of the electric field are linearly dependent on the node amplitudes $A_j$ we seek to infer since $\theta_{ij}$ is independent of $A_j$ ($\theta_{ij}$ can solely be calculated from the location of observation $i$, $\vec{r}_i$, and the SECS node location, $\vec{r}_j$). With $m$ observations of the convection electric field distributed across the analysis domain, the problem of finding the SECS node amplitudes that describe the observed convection electric field can be formulated as a set of $m$ linear equations (one per observation, index $i$) where we solve for the $n$ unknown amplitudes (index $j$).

Following Amm et al. (2010), this system of linear equations can be written in matrix form as

$$d = GA. \tag{7}$$

Here, $d$ is of size $(m, 1)$ and contains the observational data of the convection electric field in the $\hat{k}_i$ direction. If the full horizontal $\vec{E}_i$ is observed, we have information from two independent directions and it is treated as two observations in this description. If LOS observations of the plasma velocity is used, $\hat{k}_i = -\hat{k}_{\text{los},i} \times \hat{B}$ where $\hat{k}_{\text{los},i}$ is the LOS direction of plasma velocity observation $v_{\text{los},i}$ and $\hat{B}$ is a unit vector along $\vec{B}$, assumed to be vertical at high latitudes. Hence, $d_i = v_{\text{los},i} B_i$ in the $\hat{k}_i$ direction when using LOS measurements, and $B_i$ is the magnetic field strength at the location of observation $i$, $\vec{r}_i$. $A$ is a column vector of size $(n, 1)$ containing the node amplitudes. $G = G_{ij}$ is a geometry matrix of size $(m, n)$ relating the effect of a unity amplitude curl-free SECS node with pole at $\vec{r}_j$ at the location of observation $i$, $\vec{r}_i$, in direction of $\hat{k}_i$. The elements of $G$ become

$$G_{ij} = \left( \frac{1}{4\pi R_I} \cot(\theta_{ij}/2) \hat{\theta}_{ij} \right)_t \cdot \hat{k}_i \tag{8}$$

As illustrated in Figure 1a, $\theta_{ij}$ is the angle between node $j$ and the point of observation $i$. $\hat{\theta}_{ij}$ is the unit vector along the great circle connecting $\vec{r}_j$ and $\vec{r}_i$, pointing away from $\vec{r}_j$. As emphasized earlier, a coordinate conversion is needed to bring the vectors into the same coordinate system before the dot product in equation (8) can be computed, indicated by the subscript $t$. Note that the geometry matrix $G$ is solely determined by the location and LOS direction of the observations and the locations of the SECS nodes, making it independent of the observed magnitudes $d$, hence the linearity of the equations.

For many applications, and from physical considerations, we would like the solution to behave in a specific manner in specific regions. This could, for example, be a boundary condition where the plasma velocity should approach zero at some specific latitude, or that the convection should or should not cross a specific boundary, for example, into or out from the polar cap. Such constraints can be imposed by adding synthetic observations at the desired locations and in the desired direction before the inversion for $A$ in equation (7).

The exact strategy for obtaining the node amplitudes $A_j$ from the set of equations given in equation (7) depends on the particular application of the technique. As $G$ is an $(m, n)$ matrix ($m$ is number of measurements, $n$ is number of nodes), regularization will in many applications be necessary. In section 2.6 we will present details on the specific steps of the inversion used to make the plots in Figures 4 and 5, which is also identical to how the inversion is done in Paper II.

When the SECS node amplitudes $A$ have been found, the estimated curl-free electric field, $\vec{E}_{\text{SECS}}$, can in principle be evaluated at any location (except for close to the SECS nodes) by computing $G$ for the desired location(s) in the desired direction(s) and use equation (7) to compute $d$. The observation vector $d$ now represents components of $E_{\text{SECS}}$ at the desired location(s) in the desired directions, typically east and north to represent the convection electric field vector.

When applying the SECS analysis on the convection electric field across the entire high-latitude region using LOS velocity data from SuperDARN, we have found the grid displayed in Figure 2 appropriate when describing ionospheric convection features with scale sizes ≳200 km (node separation distance). This is an equal area grid that follows the MLATs, and the spatial resolution is 2° in the latitudinal direction, and the innermost ring of bins ∈ [87°,89°] MLAT is divided into four MLT bins, resulting in 480 grid cells, starting at [59°,61°] MLAT. The SECS nodes are placed at the center of the grid cells, indicated as blue dots inside the blue grid in Figure 2. As the elementary functions go to infinity at the node, one has to avoid evaluating the field too close to the node. It is beneficial to evaluate the field where the distances to the neighboring nodes are similar across the entire analysis domain. For this particular grid, this is achieved by evaluating $\vec{E}_{SECS}$ on locations that are shifted from the nodes in MLT by half of the MLT separation between nodes in the same ring. These locations are shown as the red dots in Figure 2. In this way, the grid that we evaluate for $\vec{E}_{SECS}$ reflects the level of structure that can be resolved, limited by the density of the nodes.

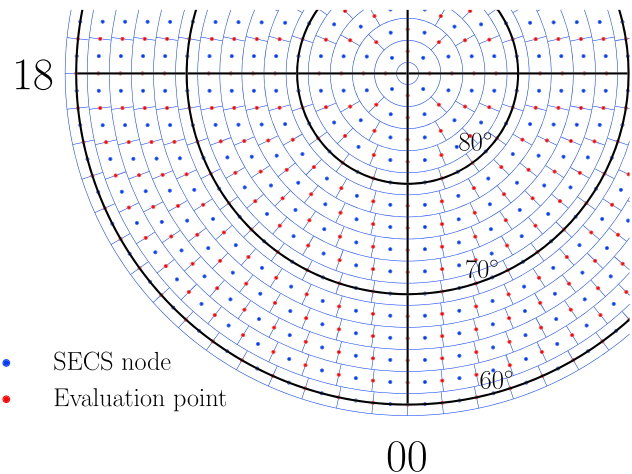

**Figure 2.** An example of a regular Spherical Elementary Convection Systems (SECS) grid (blue cells where the SECS nodes are located at the center as blue dots), here an equal area grid defined along parallels of magnetic latitude with a latitudinal width of 2°. The innermost ring of bins, [87°,89°] are separated into four magnetic local time sectors. This density of nodes are found sufficient to represent the large-scale features of the ionospheric convection electric field. Red dots illustrate locations that always have similar distance to the neighboring SECS nodes (shifted half of the magnetic local time separation in each ring of bins). These locations are found as suitable evaluation points of the electric field described by the SECS nodes.

### 2.5. A Strategy to Avoid Inverting a Very Large *G* Matrix

For the specific application of the SECS technique in Paper II, where the convection electric field is represented globally above 60° MLAT, an additional step in the preprocessing of the LOS convection data is introduced to overcome challenges with the inversion of equation (7), involving a large number of LOS observations, typically ∼ $10^6$. This strategy is to reduce the large number of LOS observations into a smaller number of average electric field vectors before inverting for *A*. We found it beneficial to again use the same grid as for the SECS nodes for this reduction, only shifted in MLT by half the MLT separation between nodes in the same MLT ring. Hence, this new grid where we compute the binned average electric field from LOS observations within the grid cell has the same center as the red dots in Figure 2 and therefore also consist of 480 cells. Due to this intermediate step, the geometry matrix *G* in equation (7) can be reduced to size (*m*, *m*). In addition, it will lead to a uniform spatial weighting (when observations are provided in every bin). The nonuniformity of LOS data coverage (observations tend to cluster) is one of the main challenges with a direct inversion for *A* from equation (7) (excluding the intermediate step of computing binned averages). Furthermore, using the binned average $\vec{E}$ as input to the inversion at the locations of the red dots in Figure 2 removes the problem of having observations very close to the SECS nodes. If using a direct inversion for *A*, a minimum allowed distance to the SECS nodes could be used, which will depend on the density of the SECS grid.

The binned average $\vec{E}$ in each grid cell is found using the same approach as described by Reistad et al. (2018), following a method outlined by Förster et al. (2008), but here applied to the convection electric field rather than the convection velocity. The average electric field vector in each grid cell, $\vec{E} = (E_{east}, E_{north})$ is found by solving the set of *m* linear equations, one per LOS observation (*i*) within the grid cell, by the method of least squares:

$$v_{los,i} B_i = (\hat{k}_{los,i} \times -\hat{B}_i) \cdot \vec{E}. \qquad (9)$$

Here, $v_{los,i}$, $\hat{k}_{los,i}$, and $\vec{B}_i$ are the LOS ionospheric plasma velocity, its associated unit vector along the LOS direction, and the magnetic field vector at the observation location, respectively. The cross product arises from the fact that we constrain the electric field in the direction perpendicular to $\hat{k}_{los,i}$ since $\vec{E} = -\vec{v} \times \vec{B}$. The LOS observations of *F* region plasma convection used in Paper II is from SuperDARN. The SuperDARN data specify $\hat{k}_{los,i}$ as degrees from local Altitude Adjusted Corrected GeoMagnetic (AACGM) coordinate system (Baker & Wing, 1989) poleward direction, positive in the eastward direction. Hence, each observation is represented in its own local coordinate system. Therefore, before equation (9) can be solved for $\vec{E}$, the observations are converted into a common frame, where $\hat{k}_{los,i}$ is now specified as degrees from the 12 MLT meridian, positive when pointing duskward. This conversion into a global coordinate system is especially important for the highest-latitude grid cells, where the equal area grid cells span several hours in MLT (6 hr in the [87°, 89°] cells). If not correcting for this, observations toward the eastern and western edges of the grid cell would lead to erroneously large squared distances when estimating $\vec{E}$, as equation (9) would then determine the distance according to the center of the grid cell. Details of the SuperDARN data processing prior to the SECS analysis described here can be found in Paper II.

### 2.6. Specific Constraints in the SECS Inversion

As emphasized in this paper, there are a number of choices needed to be made to arrive at a SECS description of the convection electric field. An optimal combination of the different ingredients can be very difficult to find and depends much on the problem at hand and the available data. In our work with representing the average ionospheric convection electric field above 60° MLAT, we have found that in particular three different quantities need to be tuned together to produce a reliable result. These quantities are the density of the grid, the degree of forcing of the solution at the low-latitude boundary, and the singular value decomposition (SVD) cutoff value used in the inversion of equation (7). For all plots shown in this and the companion paper, we use the following values, which we have found highly suitable for this particular analysis:

- *Grid*: Equal area grid with 480 cells above 60° MLAT, $\Delta$MLAT = 2°. Furthermore, we evaluate for $E_{\text{SECS}}$ on an equally dense grid, but at locations that have similar distance to the neighboring nodes, see Figure 2.
- *Boundary*: 600 synthetic observations of $E_{\text{east}} = 0$ and $E_{\text{north}} = 0$ at 59° MLAT, evenly separated in MLT. These synthetic data are displaced 1° from the 60° MLAT analysis boundary to avoid getting very close to the SECS nodes at that latitude. This weakly imposed boundary condition make the solution behave as expected toward the boundary, where observations are sparse, while having a minor influence at the polar cap latitudes.
- *SVD cutoff*: Singular values of less than 6% of the largest singular value are set to 0. In choosing a cutoff value, one has to compromise by level of spatial detail the solution can reconstruct and the amount of noise in the solution.

When these parameters are fixed, a more detailed comparison of the result of the inversion during different driving conditions can be made. We note that the exact choice of the values mentioned above does not affect the conclusions of the data analysis presented in Paper II. For the sake of reproducibility we explicitly state the values used here.

Although we use a large data set of *F* region LOS plasma velocities from SuperDARN, observations are sparse toward the low-latitude part of the analysis domain, especially on the dayside. Hence, some of the binned average $\vec{E}$ vectors in these regions are based on very few observations and are poorly defined. To reduce the importance of these grid cells in the inversion of the node amplitudes, we have introduced a weighting of the binned average $\vec{E}$ vectors found using equation (9) based on how well this region is sampled. From experimenting with data selection, we have found that the number of unique hours of observations within a grid cell is a good indicator of the data coverage. This parameter is therefore used to identify grid cells that have a potentially poorly determined $\vec{E}$. To reduce the impact of poorly determined binned average $\vec{E}$ on the solution of $A$, a weighting of the different cells based on their number of unique observational hours is applied. For grid cells having observations from more than 50 unique hours, an equal weight of 1 is applied as this is considered as good coverage. The rest of the cells (mainly below 64° MLAT on the dayside) are downweighted (in both northward and eastward direction) according to $w_j = h_j/50$ where $h_j$ is the number of unique hours cell *j* has observations from. The synthetic boundary observations are given a weight of 1. The weighting is implemented by constructing a weight matrix *w* with weights on the diagonal corresponding to how *d* in equation (7) was constructed. Then, *w* is multiplied on both sides of equation (7), before this set of equations are solved for *A* with the method of SVD.

## 3. Interpretation and Validation of the SECS Representation

In this section we discuss the physical interpretations that can be made of the inferred node amplitudes in the SECS representation of the convection electric field. We also show a comparison of the SECS representation with the corresponding convection patterns derived using spherical harmonic cap analysis on the same data set (Thomas & Shepherd, 2018) for validation purposes.

### 3.1. Interpretation of the Node Amplitudes

To get a better understanding of the assumptions made by representing $\vec{E}$ as a finite number of SECS nodes, and the physical interpretations that can be made from the node amplitudes $A_j$, we need to realize the underlying assumptions made by Amm (1997) in deriving the elementary function of the curl-free node field and its scaling to the radius and amplitude. We emphasize that the unit of $A_j$ depends on the field that is described. In our application it has units of volts, while for the original application (Amm, 1997) it has units of amperes. Since Amm (1997) described a height-integrated current field $\vec{J}_{cf}$ (A/m), the following criteria needed to be met by the elementary current field from node $j$, $\vec{J}_{cf,j}$:

$$
\begin{aligned}
&1) \; \nabla \times \vec{J}_{cf,j} \equiv 0 \\
&2) \; \text{Birkeland current entering at node } j[A] = \lim_{\theta_{ij} \to 0} \int_0^{2\pi} J_{cf,j} R_I d\phi \equiv A_j \\
&3) \; \nabla_\perp \cdot \vec{J}_{cf,j} \equiv \text{const. for} \quad \theta \neq 0 \\
&4) \; \vec{J}_{cf,j}(\theta = 180°) \equiv 0
\end{aligned}
\quad (10)
$$

1) The requirement of the current field to be curl free. 2) The Birkeland current entering at the pole, defined to be the value of $A_j$, is distributed horizontally from the pole. 3) The requirement of current continuity ensures that there is no pileup of current on the sphere. Away from the node location $\vec{r}_j$, current is then leaving the spherical surface at a constant density where const. $= \frac{-A_j}{4\pi R_I^2}$. 4) Ensures that there is no current at the point opposite to the node. These requirements uniquely define the elementary function as expressed in equation (2). From this definition, we will have the following constraints on $A_j$ from point (2) when we describe $\vec{E}_{cf}$ through equation (5):

$$
\lim_{\theta_{ij} \to 0} \int_0^{2\pi} E_j R_I d\phi = A_j. \quad (11)
$$

We can use this to relate $A_j$ to a charge density by using Gauss law and the divergence theorem around the node location:

$$
\frac{Q_j}{\epsilon_0} = \lim_{V \to 0} \int_V \nabla \cdot \vec{E}_j dV = \lim_{S \to j} \oint_S \vec{E}_j \cdot d\vec{a} = \lim_{\theta_{ij} \to 0} \int_0^{2\pi} E_j R_I d\phi \cdot h = A_j h, \quad (12)
$$

where $h$ is the height interval where $\vec{E}_j$ is given by the elementary function, needed to get a finite electric flux into/out of $S$. The closed surface $S$ we choose to be a cylindrical box of height $h$, centered at node $j$, whose radial extent is $R_I \theta_{ij} \to 0$. This surface encloses the electric charge $Q_j$. Assuming vertical magnetic field lines representing equipotentials ($E_\parallel = 0$), the magnetic field line threading node $j$ has an associated 1-D charge density, $\lambda_j$ (C/m):

$$
\lambda_j = \frac{Q_j}{h} = \epsilon_0 A_j. \quad (13)
$$

Hence, at high latitudes and above the base of the ionosphere, the nodes can be seen as infinite conducting vertical field lines with charge density $\lambda_j = A_j \epsilon_0$.

If the distance between the nodes is constant across the analysis region, as is the case in Figure 2, it is straightforward to relate the 1-D vertical charge density $\lambda_j$ to a volume charge density $\rho_j$. This can be of interest as Gauss law relates $\rho$ to $\nabla \cdot \vec{E}$. Such a relation will therefore bridge any other representation of $\vec{E}$ to corresponding node amplitudes in a SECS representation. This means that by computing $\nabla \cdot \vec{E}$ from an empirical model of $\Phi$ at the node locations in Figure 2, one can calculate the corresponding amplitudes of the curl-free nodes in a SECS representation of $\vec{E}$. Going the other way, calculating the corresponding spherical harmonic coefficients based on the SECS description is not straightforward. When approximating $\rho_j$ from $\lambda_j$, or vice versa, we assume that the 1-D charge density from the node is uniformly distributed within the area $\sigma$ defined by the uniform SECS grid cells (blue grid in Figure 2). For the grid shown in Figure 2, $\sigma = 8 \cdot 10^{10}$ m$^2$. The corresponding volume charge density then becomes $\rho_j$ (C/m$^3$) $= \lambda_j / \sigma$. By also dividing by the electron charge $e$, this can be related to the corresponding perturbation in electron density within the $j$th grid cell, $\Delta n_{e,j}$ (#/m$^3$) associated with the charged field lines, or equivalently, $\Delta n_{e,j}$ due to $\nabla \cdot \vec{E}$.

$$
\Delta n_{e,j} = \frac{\rho_j}{e} = \frac{\epsilon_0 A_j}{e\sigma} = \frac{\epsilon_0 \nabla \cdot \vec{E}}{e} \quad (14)
$$

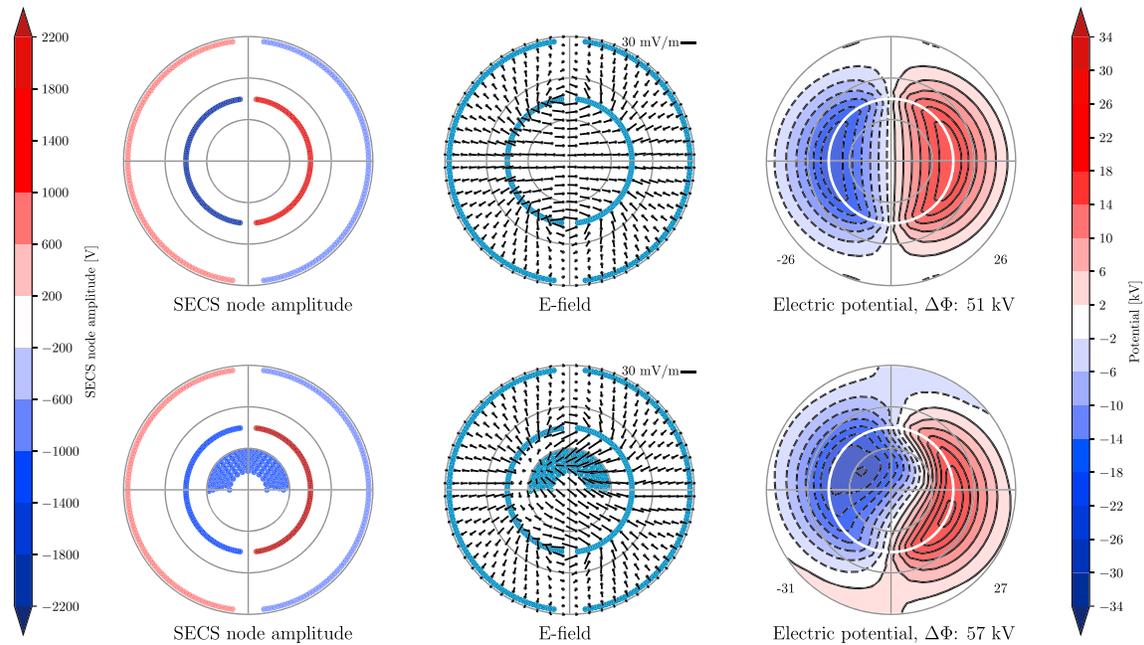

**Figure 3.** A synthetic distribution of Spherical Elementary Convection Systems (SECS) nodes (left column), its resulting $\vec{E}$ above 60° magnetic latitude (middle column), and $\Phi$ (right column). (upper row) SECS nodes placed in two circles with amplitudes chosen to reflect typical Region 1/Region 2 current morphology result in the common two-cell ionospheric convection pattern. (bottom row) An additional region of negative amplitude SECS nodes in the dayside polar cap is added. The same amount of positive amplitude is added to the Region 1 band of nodes, leaving the total charge 0. This is to mimic the influence from interplanetary magnetic field $B_y$, and correspond to adding a pattern similar to the $F_A^2$ panel in Figure 2 in Milan et al. (2015; describing the current system related to interplanetary magnetic field $B_y$) to the SECS node pattern in the upper left panel. The white circle in the rightmost panels indicate the latitude of the innermost band of Region 1 current and can be interpreted as the open/closed field line boundary.

As shown in Paper II, these values are *very* small ($\sim 10$ m$^{-3}$) compared to the $F$ region density ($\sim 10^{11}$ m$^{-3}$), hence not violating the assumption of quasi neutrality. The interpretation of $\rho = \epsilon_0 \nabla \cdot \vec{E}$, corresponding to charged vertical field lines, was also briefly mentioned by Untiedt and Baumjohann (1993, p. 292). However, we are not aware of any studies taking advantage of this representation in investigations of the ionospheric convection. Note that when expressing the charge density as in equation (14) we have not taken into account the constant divergence of $\vec{E}_j$ over the entire sphere as described in point (3) above. When applying the SECS description locally, only the spatial structures in the charge density can be resolved as the constant offset level will need to be determined from nodes all over the sphere. However, this constant contribution to the charge density is expected to be very small compared to the spatial variations across the high latitudes and is found to be $\sim 10^{-5}$ of the perturbations seen in Figure 4c when taking the sum of $A_j$ above 60° MLAT as representative of the entire globe.

### 3.2. How the Inferred SECS Node Amplitude Distribution Can Relate to Magnetospheric Drivers of Ionospheric Convection

In this subsection we show the electric field and potential from a synthetic, idealized distribution of SECS nodes. This is to get an impression of how the node amplitudes relate to the more familiar electric field and potential across the high latitudes. In Paper II we mainly focus on the interpretation of the distribution of $\Delta n_e$. Hence, this subsection as well as subsection 3.4 is of particular relevance for the interpretation of the maps in the companion paper.

The left column in Figure 3 shows two synthetic SECS node amplitude distributions above 60° MLAT in an orthogonal MLAT/MLT coordinate system. The top row has two rings, one placed at 75° MLAT and the other at 61° MLAT, with opposite sign at dawn and dusk. Each half circle band consists of 100 nodes evenly spaced in MLT. The corresponding electric field, found using equation (5) and evaluated on the same equal area grid as shown in Figure 2, is shown in the middle column as black vector pins. Here, also the locations of the SECS nodes are shown for reference (blue dots). It is evident that $\vec{E}_{SECS}$ points toward and away from the locations where SECS nodes have nonzero amplitudes. Using equation (6), we calculate the value of the electric potential $\Phi$ at the same locations as the $\vec{E}$ vectors shown in the middle panel. A contour plot of the potential values are shown in the rightmost column. The example shown in the top row of Figure 3 aims to illustrate that information about the source of the ionospheric convection can be revealed from looking at the corresponding distribution of node amplitudes that represent the convection electric field. We here demonstrate that a configuration of SECS nodes oriented in bands similar to the Region 1 and Region 2 Birkeland current pattern (Iijima & Potemra, 1978) reproduce the familiar two-cell convection pattern. We note that this convection pattern is almost identical to the idealized model of ionospheric convection as presented by Milan (2013), where the convection was modeled in response to similar bands of Region 1 and Region 2 currents as shown here.

Another example of the interpretation of the SECS node amplitude distribution is shown in the bottom row of Figure 3. As was pointed out for the symmetric two-cell pattern in the top row, the SECS node amplitudes are closely related to the Birkeland currents. This will be discussed in more detail in section 3.4. Milan et al. (2015) showed that the most common Birkeland current pattern except for the Region 1 and Region 2 bands is the one associated with interplanetary magnetic field (IMF) $B_y$. We adopt their IMF $B_y$ pattern as seen in their Figure 2, panel $F_A^2$, and add SECS nodes with a corresponding amplitude distribution on top of the SECS nodes in the top row of Figure 3. Specifically, we add 100 nodes within the dayside polar cap between 80° and 85° MLAT with negative amplitudes, corresponding to NBZ (northward $B_z$) currents during IMF $B_y$-dominated periods (Iijima et al., 1984). In addition, an amount of positive amplitude (charge) is added to every node in the innermost ring of SECS nodes at 75°, so that the sum of added IMF $B_y$ SECS node amplitudes become 0. The physical interpretation of this is that there can be no pileup of charge in the ionosphere. Hence, the upward and downward currents must on average cancel, as is the case in the elementary patterns derived by Milan et al. (2015). The sum of this "two-cell + IMF $B_y$" pattern of SECS nodes is seen in the bottom row left panel of Figure 3. The corresponding electric field and potential reveal the main large-scale features of the convection pattern during southward and $B_y$-dominated IMF, namely, a "banana-" and "orange-" shaped convection cell. Furthermore, the dusk convection cell has its minimum value inside the inner ring of SECS nodes, which can be interpreted as the open/closed field line boundary, indicated by the white circle. This is similar to how lobe reconnection during positive IMF $B_y$ affects the convection pattern in the Northern Hemisphere, leading to plasma circulation inside the polar cap in the direction shown here (Crooker & Rich, 1993). However, from looking at the electric potential only, it is not straightforward to separate and quantify the two source regions (the rings and the high-latitude region) of the convection electric field. By utilizing the technique described here, this can now be achieved by examining the estimated amplitudes of the SECS nodes in various regions.

### 3.3. Validation

In order to validate the above described methodology and demonstrate one of its applications, we investigate the ability to reproduce known variability of the electric potential with the IMF clock angle. Figure 4a shows the electric potential calculated with the above described technique for eight different orientations of the IMF clock angle based on SuperDARN LOS measurements from the same conditions as used by Thomas and Shepherd (2018) in making their Figure 5, also shown for reference in Figure 4b. These conditions are $-10° <$ tilt $< 10°$ and $1.6 < E_{sw} < 2.1$ mV/m and based on data from Northern Hemisphere only during the years 2010–2016. Furthermore, only echoes from ranges between 800 and 2,000 km are considered to reduce the likelihood of geolocation inaccuracies with multihop propagation and low-velocity $E$ region echoes. For details on the preprocessing of the SuperDARN data and selection based on IMF, see Paper II. As can be seen from Figures 4a and 4b, the potential patterns derived using these two completely different techniques agree to a high level of detail. However, we note that our values of the maximum potential difference are slightly lower than those reported by Thomas and Shepherd (2018), typically by ∼5 kV. This difference is likely related to the intermediate step of computing the binned average $\vec{E}$ before the inversion, as described in section 2.5. For a detailed comparison of the magnitudes of the binned average $\vec{E}$ versus $\vec{E}_{SECS}$, see Figure 4 in Paper II.

This comparison shows the ability of the described technique to reproduce known features of the ionospheric convection. Also shown in Figure 4c is the derived values of the SECS node amplitudes corresponding to the potential in Figure 4a. For comparison, we also show maps of the Birkeland currents corresponding to the same conditions from the Average Magnetic field and Polar current System (AMPS) model (Laundal et al., 2018). This is an empirical model of the high-latitude current system, parameterized by solar wind speed, IMF $B_y$, IMF $B_z$, dipole tilt angle, and the $F$10.7 index. When making Figure 4d we have used the values $v_{sw} = 400$ km/s, IMF $B_T = 4.5$ nT, dipole tilt $= 0°$, and $F$10.7 $= 109$ sfu. The striking similarities between the patterns of the SECS node amplitudes and Birkeland currents will be discussed in the following subsection.

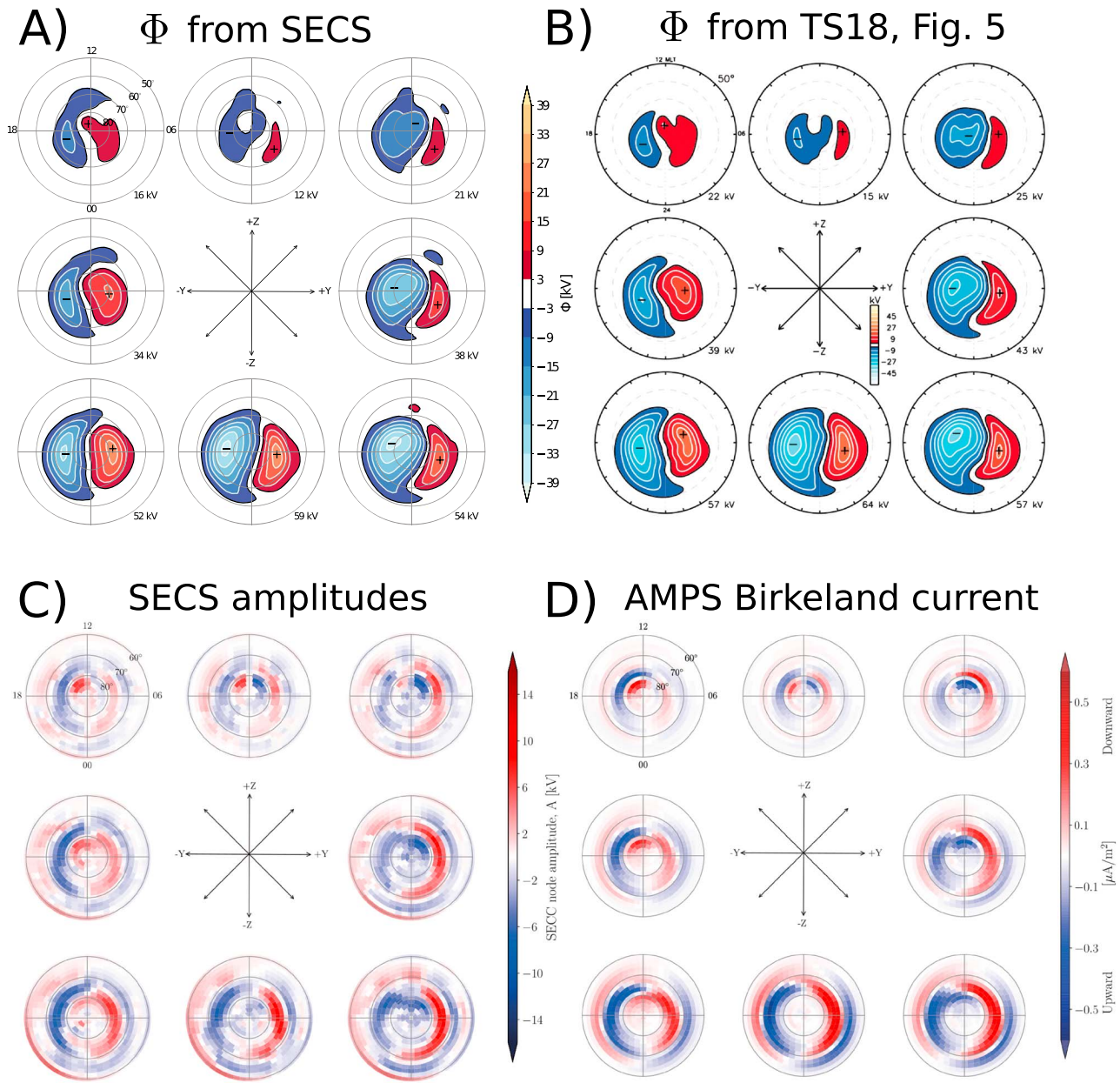

**Figure 4.** (a) Electrostatic potential calculated using the technique described in this paper using the same data selection criteria and data set as used by Thomas and Shepherd (2018) when making their Figure 5, shown here for reference in (b). The same contour spacing of 6 kV is used in (a) and (b) as well as a similar colormap for direct comparison. (c) The SECS node amplitudes that was used to make (a). (d) Statistical average pattern of Birkeland currents during the same conditions from the AMPS model (Laundal et al., 2018). AMPS = Average Magnetic field and Polar current System; SECS = Spherical Elementary Convection Systems.

### 3.4. Segmentation of the Ionospheric Convection

One of the main benefits of the SECS technique applied to ionospheric convection is the ability to separate the sources of convection. One particular application is to distinguish and quantify the contributions to the ionospheric convection from lobe and Dungey-type reconnection. In this subsection we demonstrate this ability and describe how this can be used to quantify the amount of lobe cell convection during northward IMF, which is the topic of Paper II.

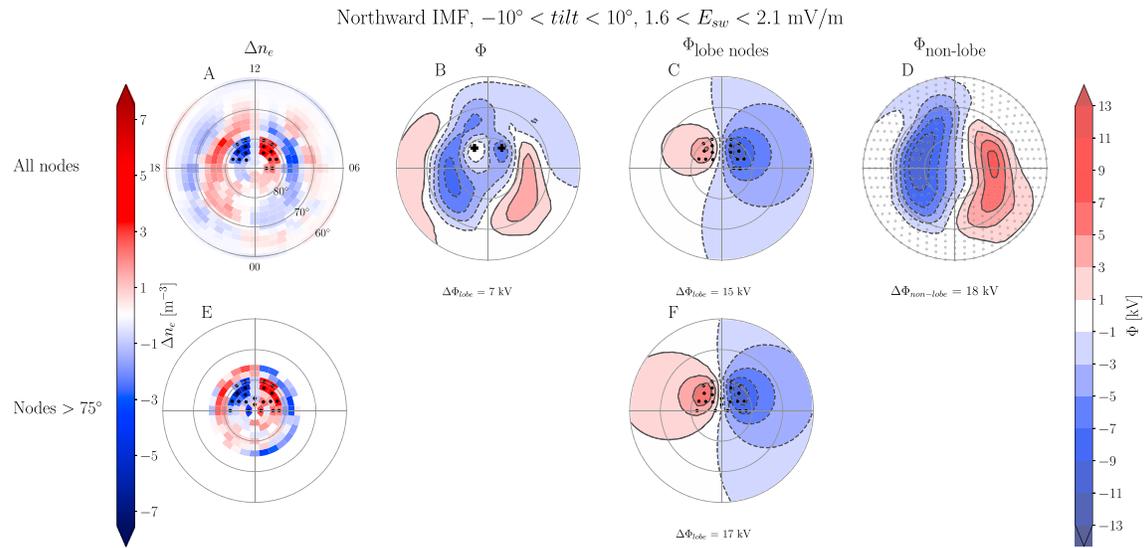

**Figure 5.** Process of identifying the sources of ionospheric convection from the maps of SECS node amplitudes. (a) The estimated SECS amplitudes converted to units of excess electrons per cubic meter using equation (14). SECS nodes within 80° magnetic latitude and between 06 and 18 magnetic local time are indicated as lobe cells and highlighted with black dots. (b) The electric potential corresoponding to (a). (c) The potential from only the nodes identified as the lobe cells in (a) (black dots). (d) The potential based only on SECS nodes outside the dayside polar cap. (e) Same as (a) but the SECS node amplitudes are estimated only within 75° magnetic latitude and based on data only from the same region. (f) Same as (c) but based on the node amplitudes from (e). IMF = interplanetary magnetic field; SECS = Spherical Elementary Convection Systems.

Figure 5 shows an example of how we identify the different sources for the ionospheric convection and the magnetic flux transport rate (potential) associated with these sources. These plots are made from SuperDARN LOS velocity measurements during northward IMF and equinox conditions, identical to how Figure 4c, middle top panel, was made.

Figure 5a shows the estimated SECS node amplitudes converted to units of excess electrons per cubic meter using equation (14), as color-filled grid cells. A pattern very similar to the average Birkeland currents during the same conditions (Figure 4d, northward IMF panel) is seen in Figure 5a. A close relationship to the field aligned currents is indeed expected. Following Milan (2013, equations (2)–(4)), current continuity, and the common decomposition of the ionospheric currents along and perpendicular to the convection electric field, one arrives at the following expression for the Birkeland current:

$$j_\| = \Sigma_P \nabla \cdot \vec{E} + \nabla \Sigma_P \cdot \vec{E} + \nabla \Sigma_H \cdot \left( \hat{B} \times \vec{E} \right). \tag{15}$$

Here, $\Sigma_H$ and $\Sigma_P$ are the height-integrated ionospheric Hall and Pedersen conductivities, respectively, and $\hat{B}$ is a unit vector along $\vec{B}$. If we assume that gradients in the Hall and Pedersen conductances are of minor importance in this regard, we can relate $j_\|$ to our estimated node amplitudes $A_j$ or $\Delta n_e$ through equation (14):

$$j_\| \approx \Sigma_P \nabla \cdot \vec{E} = \frac{\Sigma_P A_j}{\sigma} = \frac{\Sigma_P e \Delta n_e}{\epsilon_0}. \tag{16}$$

Hence, the part of the Birkeland currents not associated with conductivity gradients are directly proportional to $\Delta n_e$, or $A_j$. The close relationship seen between the panels in Figure 4c and 4d is therefore expected. We again emphasize that our patterns of $\Delta n_e$, or equivalently $\epsilon_0 \nabla \cdot \vec{E}/e$, or $\frac{\epsilon_0 A_j}{e\sigma}$, is to the first order only determined by the plasma motion and not by the ionospheric conductivity. This is in contrast to the Birkeland currents that strongly depend on the ionospheric conductivity, as shown in equation (16), and Figures 1 and 2 in Paper II. When the magnetospheric configuration can be considered stationary and there are no parallel electric fields, the large-scale ionospheric convection electric field are often considered as a "mirror image" of the coupled magnetosphere. Although some deviations from this ideal treatment may also exist in the steady state as pointed out by Hesse et al. (1997), this strong coupling allows us to relate and quantify the steady state large-scale ionospheric convection to its magnetospheric counterpart.

From the comparison of Figure 4d northward IMF panel and Figure 5a, it is easy to point out the SECS nodes that correspond to the NBZ currents seen in Figure 4d. These nodes are highlighted with a black dot in Figure 5a. In this identification we have used a threshold value of $|\Delta n_{e,j}| > 1$ m$^{-3}$ and be located at $\geqslant 80°$ MLAT and MLT $\in$ [06, 18]. In addition, we require that the sign of $\Delta n_{e,j}$ should match the expected NBZ current direction at the 80° MLAT bin to avoid selecting nodes related to the poleward edge of the Region 1 current.

The electric potential $\Phi$ associated with the SECS node amplitudes using all nodes are shown in Figure 5b. Here, two small lobe cells are seen at the same locations as where we identified the lobe cells in panel (a). The maximum and minimum locations of the lobe cells are indicated with two "+" symbols, and their potential difference is 7 kV. In addition, a two-cell pattern at slightly lower latitudes is seen. This potential difference of 7 kV can be a measure of the flux circulation in the lobe cells, but we will here argue that there is a different way to estimate this lobe potential that is more beneficial, reflecting their source region. In Figure 5c we show the potential due to the lobe SECS nodes only (black dots). This separation of the sources of the ionospheric convection is one of the benefits of representing the convection electric field as a sum of nodes, as this representation also allows for a quantification of the associated magnetic flux transport. The potential from the SECS nodes outside this dayside polar cap region is for reference shown in Figure 5d. One can see that this decomposition efficiently isolates the four-cell pattern seen in Figure 5b into the contribution from the lobe cells and the Dungey-type convection. We suggest that the potential difference related to the lobe nodes only, as shown in Figure 5c to be 15 kV, is a more realistic value of the influence from lobe reconnection, as Figure 5a demonstrates that the local value of $\Delta n_e$ appear to be closely related to $j_\parallel$ and hence a magnetospheric source, and not as much influenced by the surrounding SECS nodes. This property of $\Delta n_e$, which is mainly reflecting the influence of the local sources of convection, is investigated in Figures 5e and 5f. We here show plots in the same format as the upper row, but now $\Delta n_e$ and the potential from the identified lobe nodes are calculated from observations above 75° MLAT only. In addition, SECS nodes are only placed above 75° MLAT. Figures 5e and 5f are therefore a local variant of the same analysis that should give a similar result if the influence from distant processes is not important for the locally determined values of $\Delta n_e$. We see similar patterns and amplitudes in the dayside polar cap, and the associated potential is also similar as when including the entire high-latitude region in the inversion for $A_j$. This suggests that the value of $\Delta n_{e,j}$ reflects the local sources of influence on the ionospheric convection electric field. We therefore suggest to use the values obtained by the analysis in Figure 5c when quantifying the influence from lobe reconnection on the ionospheric convection during northward IMF, which is what we do in Paper II. If assuming a strong coupling between the ionosphere and magnetosphere ($E_\parallel = 0$ inside the polar cap), this potential difference inferred from the lobe nodes, representing the ionospheric magnetic flux transport rate around these nodes, can be interpreted as the lobe reconnection rate.

## 4. Discussion and Conclusion

This paper describes a novel technique that makes it possible to separate and quantify the different sources of ionospheric convection. The separation of the sources of lobe cell convection from the typical two-cell pattern is an obvious application of this technique, which is applied in Paper II to quantify how much the dipole tilt angle can alter the lobe cell convection.

Although very similar patterns as derived from the SECS node amplitudes (Figure 4c) can be obtained from taking the divergence of $\vec{E}$ from any representation of $\vec{E}$ (due to the relation expressed in equation (14)), the SECS representation is the most direct approach to perform this task, as the amplitudes, or equivalently the charges (see equation (13)), is what is solved for, and not a derived quantity. However, if expressing a snapshot of the global ionospheric convection, data coverage is always an issue, and additional assumptions are needed to describe the global convection. Then, including information from existing empirical models in regions without data using, for example, the map potential technique (Ruohoniemi & Baker, 1998) to represent $\Phi$, and subsequently use the relation in equation (14), we expect that a pattern reflecting the magnetospheric source of the ionospheric convection will be obtained.

The vorticity of the ionospheric convection ($\nabla \times \vec{v}$) has strong similarities to the divergence of the ionospheric convection electric field discussed in this paper and has been much used as a proxy for the Birkeland current (Sofko et al., 1995; Chisham et al., 2009; Chisham, 2017), as expressed in equation (16). We would like to point out that although there is a strikingly good correspondence between the two quantities on average (see Figure 4c vs. Figure 4d), the ionospheric convection is, at least to the first order, independent of the ionospheric conductivity. Hence, the two quantities (Birkeland current and convection) provide different information about the ionospheric electrodynamics. Taken together, they can provide information about the conductivity in the ionosphere without assuming a constant conductance. In some regions, a constant conductance might be an acceptable approximation. In Paper II we make this assumption and use equation (16) to make estimates of $\Sigma_P$ in the dayside polar cap. However, there exist more sophisticated methods like the "method of characteristics" (Amm, 2002) where the Hall and Pedersen conductances can be solved for only by assuming their ratio.

**Acknowledgments**
SuperDARN (Super Dual Auroral Radar Network) is an international collaboration involving more than 30 low-power HF radars that are operated and funded by universities and research organizations in Australia, Canada, China, France, Italy, Japan, Norway, South Africa, United Kingdom, and the United States. A large thanks to Evan Tomas from who we directly obtained the SuperDARN data. Raw files can be accessed via the SuperDARN data mirrors hosted by the British Antarctic Survey (https://www.bas.ac.uk/project/superdarn/#data) and University of Saskatchewan (https://superdarn.ca). We acknowledge the use of NASA/GSFC's Space Physics Data Facility (http://omniweb.gsfc.nasa.gov) for OMNI data. Financial support has also been provided to the authors by the Research Council of Norway under the contract 223252.